\documentclass{appolb}
\usepackage{amsmath,amssymb}
\usepackage{graphicx}

\newcommand{\be}{\begin{equation}}
\newcommand{\ee}{\end{equation}}
\newcommand{\ba}{\begin{eqnarray}}
\newcommand{\ea}{\end{eqnarray}}
\newcommand{\ban}{\begin{eqnarray*}}
\newcommand{\ean}{\end{eqnarray*}}
\newcommand \nn {\nonumber}

\begin{document}

\title{Hadron-Deuteron Correlations \\ and Production of Light Nuclei \\in Relativistic Heavy-Ion Collisions}

\author{Stanis\l aw Mr\' owczy\' nski$^{1,2}$ and Patrycja S\l o\'n$^1$
\address{$^1$Institute of Physics, Jan Kochanowski University,  
\\ ul. Uniwersytecka 7, 25-406 Kielce, Poland 
\\
$^2$National Centre for Nuclear Research, \\
ul. Pasteura 7, 02-093 Warsaw, Poland}}

\maketitle

\begin{abstract}

The production of light nuclei in relativistic heavy-ion collisions is well described by both the thermal model, where light nuclei are in equilibrium with all other hadron species present in a fireball, and by the coalescence model, where light nuclei are formed due to final state interactions after the fireball decays.  A method to falsify one of the models is proposed.  We suggest to measure a hadron-deuteron correlation function which carries information about the source of the deuterons and allows one to determine whether a deuteron is directly emitted from the fireball or if it is formed afterwards. The $K^-\!-\!D$ and $p\!-\!D$ correlation functions are computed to illustrate the statement. 

\end{abstract}

\section{Introduction}

Light nuclei are expected to be formed at the latest stage of relativistic heavy-ion collisions when the fireball decays into hadrons. Nucleons which are close to each other in phase-space still interact and can fuse into nuclei. This is the picture behind the coalescence model \cite{Butler:1963pp,Schwarzschild:1963zz}.  

The model works well in a broad range of collision energies and, in particular, it properly describes \cite{Sun:2015ulc,Sun:2017ooe,Zhu:2015voa,Zhu:2017zlb,Sombun:2018yqh} production of light (anti-)nuclei at LHC \cite{Adam:2015vda,Adam:2015yta,Acharya:2017bso}. However, the yields of light nuclei and hypernuclei \cite{Adam:2015vda,Adam:2015yta,Acharya:2017bso,Chen:2018tnh} are also accurately described by the thermodynamical model \cite{Andronic:2010qu,Cleymans:2011pe,Andronic:2017pug} with a universal temperature of 156 MeV and vanishing baryon chemical potential for all hadron species observed at midrapidity at LHC. 

This result, which has attracted a lot of interest \cite{Wang:2017smh,Mrowczynski:2016xqm,Bazak:2018hgl,Bellini:2018epz,Oliinychenko:2018ugs,Xu:2018jff,Bugaev:2018klr,Sun:2018mqq,Vovchenko:2019aoz,Cai:2019jtk}, is truly surprising as it is hard to imagine that nuclei can exist in the hot and dense environment of the fireball. The inter-particle spacing is smaller than the typical size of light nuclei and the temperature is much bigger than nuclear binding energies. Light nuclei in a fireball are thus like `snowballs in hell' \cite{PBM-2015}.

It should be stressed that the thermal and coalescence models were found long ago to predict similar yields of light nuclei \cite{DasGupta:1981xx}, and recently the result has been verified \cite{Zhu:2015voa,Mrowczynski:2016xqm} in the more advanced coalescence model \cite{Sato:1981ez,Gyulassy:1982pe,Mrowczynski:1987oid,Lyuboshits:1988yc} which properly takes into account the quantum-mechanical character of the process. 

One asks whether the final state formation of light nuclei can be quantitatively distinguished from creation in a fireball, that is, whether one of the two models can be falsified. It was suggested in \cite{Mrowczynski:2016xqm} and worked out in \cite{Bazak:2018hgl} to compare the yield of $^4{\rm He}$ to that of exotic nuclide $^4{\rm Li}$ which decays into $^3{\rm He}+p$ with the width of 6 MeV.  The alpha particle is well bound and compact while $^4{\rm Li}$ is weakly bound and loose. Since the masses are similar, the yield of $^4{\rm Li}$ is according to the thermal model about 5 times bigger than that of $^4{\rm He}$ because of 5 spin states of $^4{\rm Li}$, which has spin 2, and only one of $^4{\rm He}$. The coalescence model predicts not only a significantly smaller yield of $^4{\rm Li}$ but the yield changes with collision centrality \cite{Bazak:2018hgl}. The yield of $^4{\rm Li}$ can be experimentally obtained through a measurement of the $^3{\rm He}\!-\!p$ correlation function \cite{Pochodzalla:1987zz,Armstrong:2001mr}. The function is discussed in detail in the very recent study \cite{Bazak:2020wjn}. 

Here we present another idea how to distinguish the coalescence model from the thermal one. We suggest to measure a hadron-deuteron correlation function which carries information about the source of the deuterons and allows one to determine whether a deuteron is directly emitted from the fireball or if it is formed afterwards. We derive the hadron-deuteron correlation function treating a deuteron in Sec.~\ref{sec-hadron-proton} as in the thermal model, that is as an elementary particle emitted from a source together with all other hadrons. In Sec.~ \ref{sec-bound-D} a deuteron is treated as a neutron-proton bound state formed at the same time that the hadron-deuteron correlation is generated. 

The discussion of correlation functions is preceded in Sec.~\ref{sec-models} with a presentation of the coalescence and thermal models we repeatedly refer to. The paper is closed with a discussion of our results and conclusions.

\section{Coalescence and thermal models}
\label{sec-models}

To set the stage for further discussion we first present the coalescence and thermal models. Since we are mostly interested in deuterons we limit our consideration to the case of the simplest nuclide. We do not consider deuterons which occur as fragments of colliding nuclei but those genuinely produced at midrapidity in collider experiments at RHIC or LHC. 

\subsection{Coalescence model}
According to the coalescence model \cite{Butler:1963pp,Schwarzschild:1963zz} production of deuterons is a two step process: production of nucleons and formation of deuterons. Since the characteristic energy of the first step, which is of the order of nucleon mass,  is much bigger than that of the second step, which is of the order of binding energy, the probability to produce a deuteron with momentum ${\bf p}$ is factorized  into the probability to (independently) produce a neutron and proton with momenta ${\bf p}/2$   and the formation rate ${\cal A}$ which corresponds to a probability that nucleons fuse into the deuteron. So, one writes
\be
\label{D-n-p}
\frac{dP_D}{d^3p} = {\cal A} \, \frac{dP_n}{d^3(p/2)}  \frac{dP_p}{d^3(p/2)} ,
\ee
where $\frac{dP_i}{d^3p}$ with $i = D, n, p$ is the probability density to observe a deuteron, neutron or proton with momentum~${\bf p}$.

One usually assumes, as suggested long ago in \cite{Schwarzschild:1963zz}, that nucleons form a deuteron if they occur in a momentum sphere of a radius $p_0$. Then,
\be
\label{A-rate-pheno}
{\cal A} = \frac{3}{4} \, \frac{4 \pi p_0^3 }{3} ,
\ee
and the parameter $p_0$, which is roughly a deuteron internal momentum, is a free parameter of the model to be inferred from experimental data. The nucleons are assumed to be unpolarized and the spin factor $3/4$ takes into account that there are 3 spin states of a spin-one deuteron and 4 spin states of a nucleon pair. 

It is also often required that nucleons, which fuse into a deuteron, must be close to each other not only in the momentum space but in the coordinate space as well, see e.g. \cite{Baltz:1993jh}. The formula (\ref{A-rate-pheno}) is then modified. 

In the relativistically covariant coalescence model one uses the Lorentz invariant nucleon momentum distributions in the relation analogous to (\ref{D-n-p}) and modifies the coalescence rate formula (\ref{A-rate-pheno}) accordingly, see e.g. \cite{Sato:1981ez,Mrowczynski:1987oid}. To avoid the complication we consider the deuteron formation in the center-of-mass frame of the neutron-proton pair where the process can be treated nonrelativistically even so momenta of nucleons are relativistic in both the rest frame of the source and in the laboratory frame. The point is that the formation rate is non-negligible only for small relative momenta of the nucleons. 

It should be stressed that the phenomenological approaches to production of light nuclei, which are based on the formulas (\ref{D-n-p}) and (\ref{A-rate-pheno}) or their variations, do not take into account a quantum-mechanical character of the process of a bound state formation. However, it was discovered by Sato and Yazaki  \cite{Sato:1981ez} and discussed later on by several authors, see e.g. \cite{Gyulassy:1982pe,Mrowczynski:1987oid,Lyuboshits:1988yc}, that the formation of a nucleus driven by final state interactions is fully analogous to the process responsible for short range correlations observed among final state hadrons. Therefore, the quantum-mechanical formula which gives the deuteron formation rate is almost identical to that of neutron-proton correlation function \cite{Mrowczynski:1992gc}. Thus, it reads \cite{Sato:1981ez}
\be
\label{D-rate}
\mathcal{A} = \frac{3}{4} (2\pi)^3 \int d^3r_n \, d^3 r_p \, 
D({\bf r}_p) \, D({\bf r}_n) |\psi_D({\bf r}_n, {\bf r}_p)|^2 ,
\ee
where the source function $D({\bf r})$ is the normalized probability distribution of emission points and $\psi_D({\bf r}_n, {\bf r}_p)$ is the deuteron wave function. 

The formula  (\ref{D-rate}) is written as for the instantaneous emission of the neutron and proton but the time duration of the emission process can be easily incorporated \cite{Koonin:1977fh}. A possible momentum dependence of the source function and other aspects of the formula (\ref{D-rate}) are discussed in more detail in Sec.~\ref{sec-hadron-proton} in the context of two-particle correlation function. 

We close the presentation of the coalescence model by saying that whenever we refer to the model we keep in mind the expression (\ref{D-n-p}) with the formation rate given by Eq.~(\ref{D-rate}). 

\subsection{Thermal model}

The fundamental postulate of the thermal model is the equipartition of fireball's energy among all degrees of freedom of the system. Therefore, light nuclei are assumed to be populated as all other hadrons and when the fireball decays the nuclei show up in a collision final state. Their yield reflects a thermodynamic state of the fireball at the moment of chemical freeze-out when inelastic collisions of fireball's constituents become no longer operative.

A microscopic mechanism responsible for production of light nuclei in the fireball is unspecified and may be even unknown. Since the temperature is much bigger than the nuclear binding energies and the inter-particle spacing is smaller than the typical size of light nuclei, it is hard to imagine that the nuclei can actually exist in the fireball. Therefore, proponents of the thermal model argue \cite{Andronic:2017pug} that the final state nuclei originate from compact colorless objects of quarks and gluons with quantum numbers of light nuclei. These compact objects are suggested to be present in the fireball together with all other hadrons. 

\subsection{Do the models differ?}

One wonders whether the production mechanisms of light nuclei behind the coalescence and thermal models are physically different from each other. The coalescence is a microscopic picture while the direct thermal production is a macroscopic description. 

One can argue that instead of the two models we should rather consider, as in the study \cite{Oliinychenko:2018ugs}, hadron-hadron and hadron-deuteron interactions which are responsible for a deuteron production and disintegration in a fireball before its decay. Such an approach is physically sound if a particle source and an average inter-hadron spacing in the source are both much bigger than a deuteron size. Additionally the lifetime of the source should be much longer than the characteristic time of deuteron formation. 

However, the assumptions are rather far from reality of relativistic heavy-ion collisions. The deuteron radius is about 2 fm and the time of deuteron formation, which is of the order of the inverse binding energy, is roughly 100 fm/$c$. The size of the particle source is of the same order as the deuteron radius, the inter-hadron spacing in the source is smaller than a deuteron, and the lifetime of the source is significantly shorter than the deuteron formation time. 

The coalescence mechanisms of deuteron formation and direct thermal production are physically different in relativistic heavy-ion collisions because the particle source is small and dense when compared to a deuteron and the source lifetime is shorter than the deuteron formation time. According to the coalescence model, light nuclei are formed long after nucleons are emitted from the source. The thermal model assumes that light nuclei are emitted directly from the source.

\section{hadron-proton correlation function}
\label{sec-hadron-proton}

We start a discussion of correlation functions with the hadron-proton correlation function. The hadron will be identified with either a negative kaon or a proton. The $h\!-\!p$ correlation function $\mathcal{R}$ is defined as
\be
\frac{dP_{h p}}{d^3p_h  d^3p_p} = \mathcal{R}({\bf p}_h , {\bf p}_p) \, \frac{dP_h}{d^3p_h}  \frac{dP_p}{d^3p_p},
\ee
where $\frac{dP_h}{d^3p_h}$, $\frac{dP_p}{d^3p_p}$ and $\frac{dP_{h p}}{d^3p_h d^3p_p}$ are probability densities to observe $h$, $p$ and $h\!-\!p$ pairs with momenta ${\bf p}_h$, ${\bf p}_p$ and $({\bf p}_h, {\bf p}_p)$. 
If the correlation results from quantum statistics and/or final state interactions, the correlation function is known to be \cite{Koonin:1977fh,Lednicky:1981su}
\be
\label{def-fun-cor}
\mathcal{R}({\bf p}_h , {\bf p}_p) = \int d^3 r_h \, d^3 r_p \, 
D({\bf r}_h) \, D({\bf r}_p) |\psi ({\bf r}_h,{\bf r}_p)|^2 ,
\ee
where the source function $D({\bf r})$ is, as previously, the probability distribution of emission points and $\psi({\bf r}_h,{\bf r}_p)$ is the wave function of the hadron and proton in a scattering state. 

Before we discuss the femtoscopic formula (\ref{def-fun-cor}) in more detail, let us eliminate the center-of-mass motion of the $h\!-\!p$ pair in a non-relativistic manner. We introduce the center-of-mass variables 
\be
\label{CM-2-variables}
{\bf R} \equiv \frac{m_h {\bf r}_h + m_p{\bf r}_p }{M} ,
~~~~~~
{\bf r}_{h p} \equiv  {\bf r}_h - {\bf r}_p ,
\ee
where $M \equiv m_h + m_p$, and we write down the wave function as $\psi_{\bf q}({\bf r}_h,{\bf r}_p) = e^{i {\bf R} {\bf P}}\phi ({\bf r}_{h p})$ with ${\bf P}$ and ${\bf q}$ being the momentum of the center of mass and the momentum in the center-of-mass frame of the hadron-proton system. The correlation function (\ref{def-fun-cor}) is then found to be
\be
\label{fun-corr-relative}
\mathcal{R}({\bf q}) = \int d^3 r_{hp} \, D_r ({\bf r}_{h p}) |\phi_{\bf q}({\bf r}_{h p})|^2 ,
\ee
where the `relative' source is 
\be
\label{D-r-source-def}
D_r({\bf r}_{h p}) \equiv \int d^3 R \, D \Big ({\bf R} + \frac{m_p}{M}{\bf r}_{h p} \Big) 
\, D \Big ({\bf R}-\frac{m_h}{M} {\bf r}_{h p} \Big ) .
\ee

Let us discuss the formula (\ref{fun-corr-relative}) which will be used to compute correlation functions. 

\subsection{Reference frame}

We consider the $h\!-\!p$ correlations, as the deuteron formation, in the center-of-mass frame of the pair and we treat the formula (\ref{def-fun-cor}), similarly as (\ref{D-rate}), as nonrelativistic even though the hadron and proton momenta are typically relativistic in both the rest frame of the source and in the laboratory frame. A relativistic description of strongly interacting particles faces difficulties particularly severe when bound states like deuterons are involved. The correlation function, however, significantly differs from unity only for small relative momenta. Therefore, the relative motion can be treated as nonrelativistic and the corresponding wave function is a solution of the Schr\"odinger equation. The source function, which is usually defined in the source rest frame, needs to be transformed to the center-of-mass frame of the pair as discussed in great detail in \cite{Maj:2009ue}. 

\subsection{Source function}
\label{sec-source-fun}

We assume that the source function is time-independent and consequently the formula  (\ref{def-fun-cor}) is written as for the instantaneous emission of the two particles. The time duration of the emission process can be easily taken into account \cite{Koonin:1977fh} but if one uses an isotropic Gaussian source, as we do for the reasons explained below, the time duration $\tau$ simply enlarges the effective radius of the source from $R_s$ to $\sqrt{R_s^2 + v^2\tau^2}$ where $v$ is the velocity of the particle pair relative to the source. 
 
In general a single-particle source function is time dependent and anisotropic \cite{Shapoval:2013bga} and to disentangle temporal and different spatial sizes of the source one needs a precise measurement of correlation functions. This is easily achieved in case of pions or kaons, but is difficult for particles which are not so abundantly produced. In this case one uses, see {\it e.g.} \cite{Acharya:2019ldv}, the isotropic Gaussian source 
\be
\label{D-Gauss}
D ({\bf r}) = \Big({\frac{1}{2 \pi R_s^2}}\Big)^{3/2} e^{ -\frac{{\bf r}^2}{2R_s^2} } ,
\ee
with $\sqrt{3}R_s$ being the root-mean-square effective radius of the source. 

As already mentioned, the source function (\ref{D-Gauss}) should be transformed to the rest frame of the $h\!-\!p$ pair. This transformation makes the source anisotropic because the effective radius along the pair velocity is elongated, not contracted, as one can naively expect, see \cite{Maj:2009ue} for details. However, if the correlation function is averaged over the direction of ${\bf q}$, as done when the statistics of correlated pairs is not high enough, we deal with the isotropic source with $R_s$ being the effective radius which combines the temporal and spatial sizes of the source. 

The Gaussian parametrization of the source function (\ref{D-Gauss}) is not only convenient for analytical calculations but there is an empirical argument in favor of this choice. The imaging technique \cite{Brown:1997ku} allows one to infer the source function from a two-particle correlation function provided the inter-particle interaction is known. The technique applied to experimental data from relativistic heavy-ion collisions showed that non-Gaussian contributions to the source functions are rather small and do not much influence the correlation functions \cite{Alt:2008aa}.    

With the Gaussian single-particle source function (\ref{D-Gauss}), the relative source (\ref{D-r-source-def}) equals
\be
\label{D-r-Gauss}
D_r ({\bf r}) = \Big({\frac{1}{4 \pi R_s^2}}\Big)^{3/2} e^{ -\frac{{\bf r}^2}{4R_s^2} } ,
\ee  
which is independent of particle masses even so the variable ${\bf R}$ given by Eq.~(\ref{CM-2-variables}) depends on $m_h$ and $m_p$. 

The single particle source function (\ref{D-Gauss}) is assumed to be independent of particle's momentum and particle's mass. This is not quite right as, in general, a source radius depends on both particle's mass $m$ and momentum. More precisely, it scales with the particle's transverse mass $m_\perp \equiv \sqrt{m^2 + p_\perp^2}$.  For the case of one-dimensional analysis relevant for our study, the effect is well seen in Fig.~8. of Ref.~\cite{Adam:2015vja} where experimental data on Pb-Pb collisions at LHC, which are of particular interest for us, are presented. The dependence of the source radius on $m_\perp$ is evident when we deal with pions and $m_\perp \lesssim 0.9~{\rm GeV}$. However, the dependence becomes much weaker for protons when $m_\perp \gtrsim 1.0~{\rm GeV}$. The figure shows that the radius of proton source tends to decrease in central Pb-Pb  collisions when $m_\perp $ grows from 1.1 GeV to 1.7 GeV but the decrease is not seen for the collision centrality $10-30\%$ nor $30-50\%$. The behavior is well understood as the decrease of the source radius with growing $m_\perp $ is caused by the collective radial flow which is stronger in central than in peripheral collisions. 

In case of proton-deuteron correlations, which are discussed Sec.~\ref{sec-p-D-corr} and play a key role in our proposal, the interval of $m_\perp $ from 1 to 2 GeV is of crucial importance. The experimental data from non-central collisions, which are presented in Fig.~8  of Ref.~\cite{Adam:2015vja}, show no dependence of the source radius on $m_\perp $ in the interval. Since we are interested in rather peripheral collisions, where the source radii are sufficiently small and the effect we suggest to measure is significant, it is legitimate to assume that the source radius is independent of particle's transverse mass. 

The assumption can be relaxed but the relative source function (\ref{D-r-source-def}) becomes rather complicated. We intend to quantitatively study the effect of transverse-mass dependence of source radii in future but for now we keep the source function (\ref{D-Gauss}) independent of the mass. Therefore, it is the same for protons, kaons and deuterons. In case of kaons the assumption is not fulfilled but, as we show in Sec.~\ref{sec-K-D-corr}, the kaons are anyway not useful for our proposal.

\subsection{Wave function and Coulomb interaction}

If the Coulomb interaction is absent but there is a short-range strong interaction, the wave function can be chosen, as proposed in \cite{Lednicky:1981su}, in the asymptotic scattering form
\be 
\label{scatt-wave-fun}
\phi_{\bf q}({\bf r}) = e^{iqz}+f(q)\frac{e^{iqr}}{r} ,
\ee
where $q \equiv |{\bf q}|$ and $f(q)$ is the $s-$wave (isotropic) scattering amplitude. 

With the source function (\ref{D-r-Gauss}) and the wave function (\ref{scatt-wave-fun}), the correlation function (\ref{fun-corr-relative}) equals
\ba
\nn
\label{fun-corr-final}
&& \mathcal{R}(q) 
= 1 + \frac{1}{2R_s^2}|f(q)|^2 - \frac{1 -  e^{-4R_s^2 q^2}}{2R_s^2 q} \, \Im f(q) 
\\[2mm]
&+& \frac{1}{2 \pi^{1/2}R_s^3 q} \, \Re f(q)  \int^\infty_0 dr \,e^{-\frac{r^2}{4R_s^2}} \sin(2qr) .
\ea
The remaining integral needs to be taken numerically.  The formula (\ref{fun-corr-final}) has been repeatedly used to compute correlation functions of various two-particle systems. 

When one deals with charged particles, the formula (\ref{scatt-wave-fun}) needs to be modified because the long-range electrostatic interaction influences both the incoming and outgoing waves. However, the Coulomb effect can be approximately taken into account \cite{Gmitro:1986ay} by multiplying the correlation function by the Gamow factor that equals
\be 
\label{Gamow}
G(q) = \pm {2 \pi \over a_B q} \,
{1 \over {\rm exp}\big(\pm {2 \pi \over a_B q}\big) - 1} ,
\ee
where the sing $+$ ($-$) is for the repelling (attracting) particles and $a_B$ is the Bohr radius of the pair. 

If we treat a deuteron as an elementary particle, the formula (\ref{fun-corr-final}) with the Gamow factor (\ref{Gamow}) can be used to compute the $h\!-\!D$ correlation function.

\subsection{Reliability of femtoscopic formula}

The femtoscopic formula (\ref{def-fun-cor}), which is critically discussed in the review article \cite{Lisa:2005dd}, is simple but it is well justified in a midrapidity domain of heavy-ion collisions at RHIC or LHC. First of all, a momentum scale of hadron production is much bigger than a characteristic momentum scale of inter-hadron femtoscopic correlation. Consequently, the cross section to produce a pair of correlated particles can be factorized into the cross section to produce a pair of mutually independent particles and the correlation function of the two particles determined by quantum statistics and/or final state interactions of particles of interest. 

The second key circumstance is that the hadronic matter of the fireball at freeze-out is in a thermodynamic equilibrium as it follows from an evident success of the thermal and hydrodynamic models in describing experimental data, see {\it e.g.} \cite{Andronic:2017pug}. Therefore, a density matrix of the system is essentially diagonal or equivalently the random phase approximation  is applicable. Consequently, the system can be described in terms of probabilities not amplitudes. 

Further on, the thermal model shows that the hadron gas in thermal equilibrium can be treated as a mixture of classical ideal gases of hadrons of different species. Therefore, inter-hadron correlations in a fireball are rather weak and the many-body density matrix can be factorized into a product of single particle matrices. For this reason, the single-particle source functions enter the femtoscopic formula (\ref{def-fun-cor}).

Although the femtoscopic formula is well justified, as explained above, it is still phenomenological and it is difficult to quantify its accuracy. However, there were performed some consistency tests which show that the formula is indeed accurate. In particular, the source parameters obtained for pairs of identical charged kaons and of neutral kaons agree very well with each other even so the inter-particle interactions are rather different \cite{Adam:2015vja}. It was also shown \cite{Adamczyk:2015hza} that the parameters of antiproton-antiproton scattering extracted for the $\bar{p}\!-\!\bar{p}$ correlation function agree with the well-known parameters of the proton-proton scattering. This must be the case as long as the matter-antimatter symmetry holds.

\section{Hadron-deuteron correlation function}
\label{sec-bound-D}

We derive here the $h\!-\!D$ correlation function treating the deuteron as a neutron-proton bound state created due to final state interactions similarly to the $h\!-\!D$ correlation. Then, the correlation function is defined as
\be
\frac{dP_{h D}}{d^3p_h \, d^3p_D} = \mathcal{R}({\bf p}_h , {\bf p}_D) \mathcal{A} \, 
\frac{dP_h}{d^3p_h} \frac{dP_n}{d^3p_n} \frac{dP_p}{d^3p_p} ,
\ee
where ${\bf p}_n = {\bf p}_p = {\bf p}_D/2$. The deuteron formation rate $\mathcal{A}$, which is defined by Eq.~(\ref{D-n-p}), is given by the formula (\ref{D-rate}). 

The correlation function multiplied by the deuteron formation rate equals
\ba
\nn
\mathcal{R}({\bf p}_h , {\bf p}_D) \, \mathcal{A} 
&=& \frac{3}{4} (2 \pi )^3 \int d^3 r_h \, d^3 r_n \, d^3 r_p \, D({\bf r}_n) \, D({\bf r}_p) 
\\[2mm] 
\label{fun-corr-pi-D-A}
&& \times D({\bf r}_h) 
|\psi_{h n p}({\bf r}_h, {\bf r}_n, {\bf r}_p)|^2 ,
\ea
where $\psi_{h n p}({\bf r}_h, {\bf r}_n, {\bf r}_p)$ is the wave function of a $h\!-\!D$ system. The spin factor $3/4$ has the same origin as that in Eq.~(\ref{D-rate}). 

Using the center-of-mass variables analogous to (\ref{CM-2-variables}), the deuteron formation rate (\ref{D-rate}) is found as
\be
\label{D-rate-relative}
\mathcal{A} 
= \frac{3}{4} (2 \pi)^3 \int d^3r_{np} \, D_r ({\bf r}_{np}) |\varphi_D({\bf r}_{np})|^2 ,
\ee
where $D_r ({\bf r}_{np})$ is the `relative' source (\ref{D-r-source-def}) and $\varphi_D({\bf r}_{np})$ is the deuteron wave function of relative motion.

\begin{figure}[t]
\centering
\includegraphics[scale=0.29]{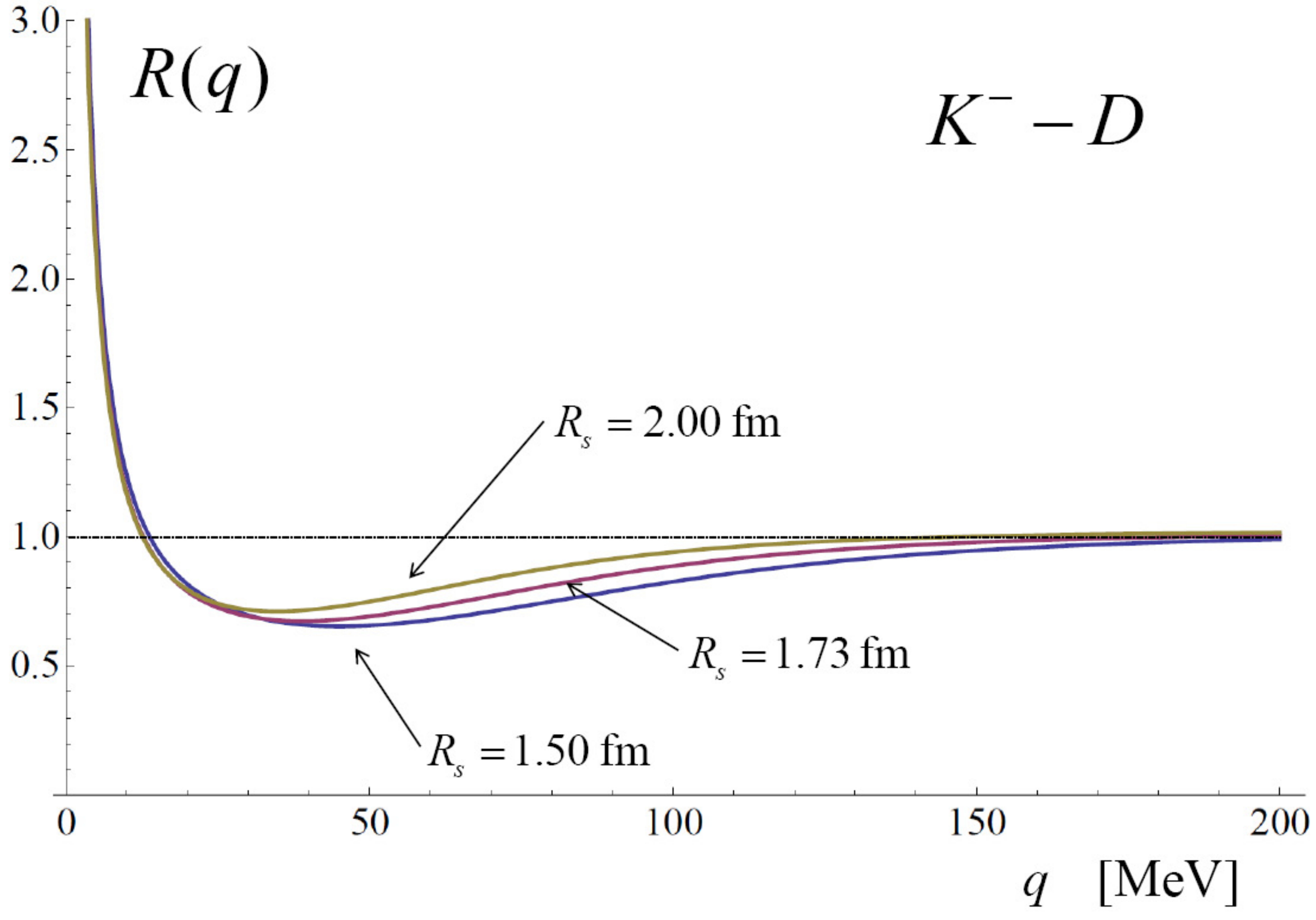}
\vspace{-2mm}
\caption{$K^-\!\!-\!D$ correlation function}
\label{fig-Kminus-D}
\end{figure}

To compute the correlation function (\ref{fun-corr-pi-D-A}), we introduce the Jacobi variables of a three-particle system 
\be 
\label{Jacobi-3}
\left\{ \begin{array}{ll}
{\bf R}\equiv \frac{m_n{\bf r}_n + m_p{\bf r}_p + m_h {\bf r }_h}{M} ,
\\[1mm]
{\bf r}_{np} \equiv  {\bf r}_n-{\bf r}_p ,
\\[1mm]
{\bf r}_{h D} \equiv  {\bf r}_h - \frac{m_n {\bf r}_n + m_p {\bf r}_p}{m_D} ,
\end{array} \right.
\ee
with $M \equiv m_n + m_p +m_h$, $m_D \equiv m_n + m_p$ and we write down the wave function as
\be
\label{wave-fun-hnp}
\psi_{h n p}({\bf r}_h, {\bf r}_n, {\bf r}_p) = e^{i{\bf P}{\bf R}} \, \psi_{h D}^{\bf q}({\bf r}_{h D}) \, 
\varphi_D ({\bf r}_{np}) .
\ee

Using the Gaussian source (\ref{D-Gauss}), the integral over the center-of-mass position ${\bf R}$  in Eq.~(\ref {fun-corr-pi-D-A}) gives
\be
\label{relative-sources-3}
\int d^3 R \,D ({\bf r}_n) \, D({\bf r}_p) \, D({\bf r}_h)
=   D_r({\bf r}_{np})  \, D_{3r}({\bf r}_{h D}) ,
\ee
where $ D_r({\bf r})$ is again given by Eq.~(\ref{D-r-Gauss}) and the normalized function $ D_{3r}({\bf r})$ equals
\be
\label{source-r-pi-D}
\mathcal{D}_{3r}({\bf r}) = \Big(\frac{1}{3 \pi R_s^2} \Big)^{3/2} e^{-\frac{{\bf r}^2}{3R_s^2}} .
\ee

As a result of the integration over ${\bf R}$ in the right-hand-side of Eq.~(\ref{fun-corr-pi-D-A}), the formation rate (\ref{D-rate-relative}) factors out. Consequently, the rate, which is also present in the left-hand-side of Eq.~(\ref{fun-corr-pi-D-A}), drops out and the correlation function equals
\be
\label{fun-corr-pi-D-bound}
\mathcal{R}({\bf q}) = \int d^3 r_{h D} \, 
D_{3r}({\bf r}_{h D}) \, |\psi_{h D}^{\bf q} ({\bf r}_{h D})|^2.
\ee

The formula (\ref{fun-corr-pi-D-bound}) has the same form as (\ref{fun-corr-relative}) but the source function differs. When deuterons are directly emitted from the fireball as `elementary' particles the radius of deuteron source is the same as the radius of proton source. When deuterons are formed only after emission of nucleons from the fireball, the source becomes bigger because the  deuteron formation is a process of spatial extent. More quantitatively, the source radius of deuterons treated as bound states is bigger by the factor $\sqrt{4/3}\approx 1.15$ than that of `elementary' deuterons. 

\subsection{$K^-\!-\!D$ correlation function}
\label{sec-K-D-corr}

To see how sensitive the correlation functions (\ref{fun-corr-relative}) and (\ref{fun-corr-pi-D-bound}) are to the source radius, we first consider the $K^-\!\!-\!D$ system which is under study by the ALICE Collaboration \cite{Graczykowski-Janik-Kisiel}. The $s-$wave amplitude is taken as 
\be
\label{ampli}
f(q) = - \frac{a}{1 + i qa},
\ee 
where the scattering length $a$ is $(1.46 - 1.08 i) \; {\rm fm}$ \cite{Doring:2011xc}. The length is complex because there are open inelastic channels of $K^-\!\!-\!D$ scattering even at $q=0$. The $K^-\!\!-\!D$ correlation function, which is computed using the formula (\ref{fun-corr-relative}) together with the Gamow factor (\ref{Gamow}), is shown in Fig.~\ref{fig-Kminus-D} for three values of $R_s$ such that $R_s = 2.00 \; {\rm fm} = \sqrt{\frac{4}{3}}\cdot 1.73 \; {\rm fm} = \frac{4}{3}\cdot 1.50 \; {\rm fm}$. 

The scenario of deuterons directly emitted from the fireball and that of deuterons formed due to final state interactions correspond to the two neighboring curves in Fig.~\ref{fig-Kminus-D}. The curves are close to each other and thus it would be very difficult to distinguish the two scenarios because of both experimental and theoretical uncertainties. The $K^-\!\!-\!D$ system is not well suited for the purpose because it is not very sensitive to $R_s$ and the sensitivity even drops when $R_s$ grows. 

The  $h\!-\!D$ correlation function is shaped by Coulomb and strong interactions. The effect of Coulomb interaction is almost independent of the source radius, as long as the radius is much smaller than the Bohr radius. If the Gamow factor is applied to take into account the Coulomb interaction, the effect is fully independent of $R_s$. Since $a_B \gg R_s$ in high-energy nucleus-nucleus collisions, the correlation function dominated by the Coulomb interaction only weakly depends on $R_s$. The $h\!-\!D$ correlation function depends on $R_s$ mostly due to strong interactions. Therefore, one should choose a system where the strong interaction is truly strong to get a correlation function sensitive to the source radius. The best choice seems to be a proton-deuteron pair. 

\subsection{$p\!-\!D$ correlation function}
\label{sec-p-D-corr}

In case of $p\!-\!D$ system the Coulomb effect is of opposite sign to that in $K^-\!\!-\!D$ and the effect of strong interactions is stronger. Since the $p\!-\!D$ pair can have spin 1/2 or 3/2 there are two interaction channels. The $s-$wave scattering lengths of $p\!-\!D$ scattering in the spin 1/2 and 3/2 channels are, respectively, 4.0 fm and 11.0 fm \cite{Black:1999duc}. Since nucleons are assumed to be unpolarized the $p\!-\!D$ correlation function is computed as the average  
\be
\mathcal{R}({\bf q}) = \frac{1}{3}\, \mathcal{R}^{1/2}({\bf q}) + \frac{2}{3}\, \mathcal{R}^{3/2}({\bf q}) ,
\ee
where the weights factors 1/3 and 2/3 reflect the numbers of spin states in the two channels. 

\begin{figure}[t]
\centering
\includegraphics[scale=0.29]{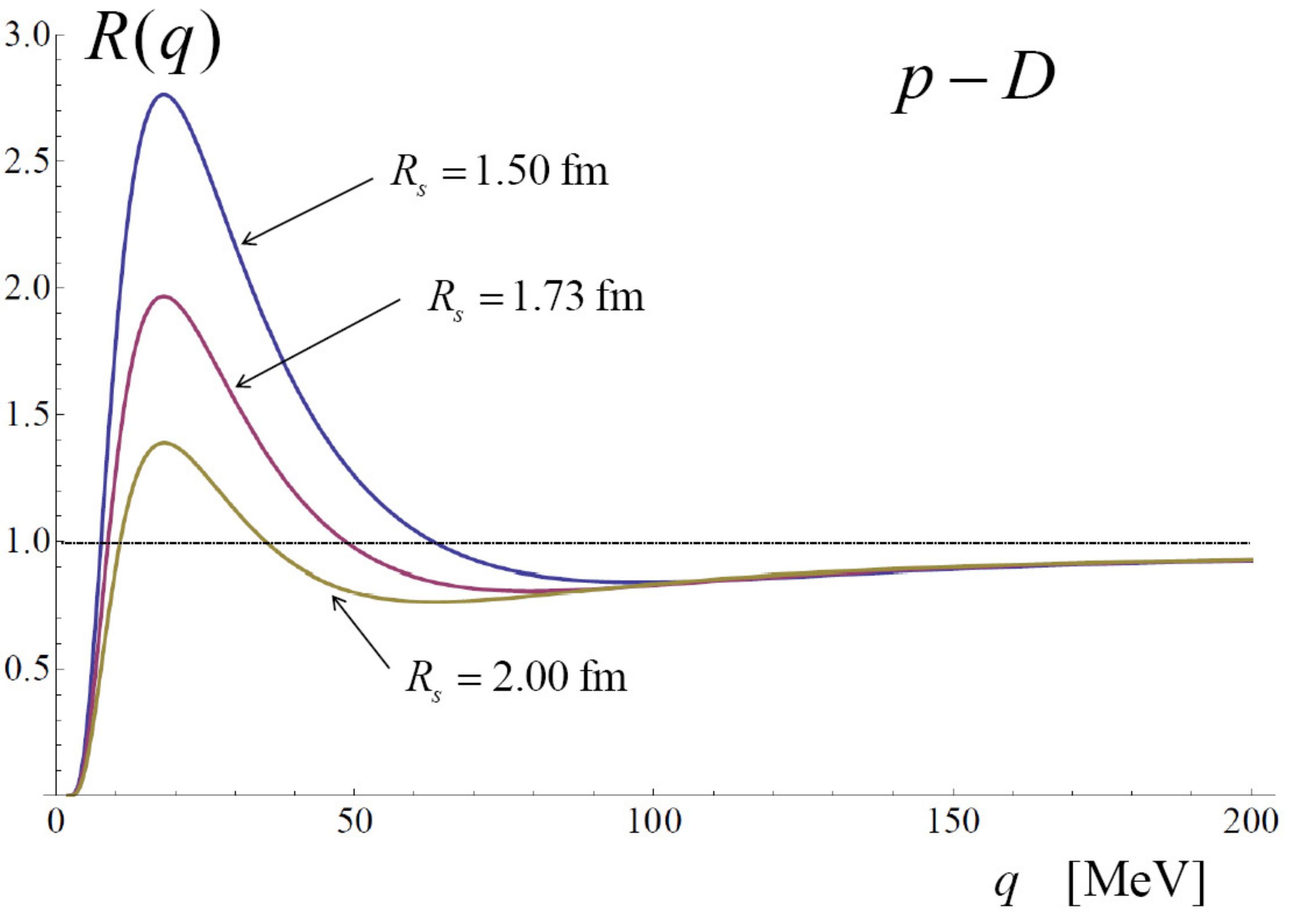}
\vspace{-2mm}
\caption{$p\!-\!D$ correlation function}
\label{fig-p-D}
\end{figure}

The average $p\!-\!D$ correlation function is shown in Fig.~\ref{fig-p-D} for three values of the source radius. As previously the values are such that $R_s = 2.00 \; {\rm fm} = \sqrt{\frac{4}{3}}\cdot 1.73 \; {\rm fm} = \frac{4}{3}\cdot 1.50 \; {\rm fm}$. The function strongly depends on $R_s$. Therefore, it should be possible to infer the source radius from experimentally measured  $p\!-\!D$ function and compare it to $R_s$ obtained from the $p\!-\!p$ correlation function. If deuterons are directly emitted from the fireball, the radii of proton and deuteron sources are the same. If deuterons are formed due to final state interactions, the radius of deuteron source is bigger by the factor $\sqrt{4/3}$. 

The dependence of the $p\!-\!D$ correlation function on $R_s$  becomes weaker as $R_s$ grows. Consequently, the analysis of higher $p_T$ particles from non-central events, when the sources are relatively small, is preferred. 

\section{Discussion and conclusions}

Our proposal to distinguish the scenario of deuterons directly emitted from the fireball from that of deuterons formed due to final state interactions does not relay on an absolute value of the source size inferred from the $p\!-\!D$ correlation function but on a comparison of source size parameters inferred from the $p\!-\!D$ and $p\!-\!p$ correlation functions. Therefore, systematic uncertainties of the femtoscopic method, both experimental and theoretical, are not of crucial importance here, as they are expected to influence in a similar way the source parameters inferred from the $p\!-\!D$ and $p\!-\!p$ correlation functions. 

We note that the size of the proton source in pp collisions at LHC was measured with an experimental accuracy of 7\% where the statistical error is only 2\% \cite{Acharya:2018gyz}. Our proposal requires an accuracy better than 15\% which, however, does not include systematic experimental and theoretical uncertainties. Therefore, the required accuracy of the measurement seems achievable. 

We have shown that a hadron-deuteron correlation function carries information about the source of the deuterons and allows one to determine whether a deuteron is directly emitted from the fireball or if it is formed afterwards. The $K^-\!\!-\!D$ correlation function is not well suited for our purpose because it weakly depends on the source radius. The $p\!-\!D$ correlation function is a better choice as the effect of strong interactions in the $p\!-\!D$ system is more pronounced and the correlation function is more sensitive to $R_s$.

We recommend a simultaneous measurement of $p\!-\!p$ and $p\!-\!D$ correlation functions. The former, which has been repeatedly measured, can be used to obtain the radius of the nucleon source and the latter would determine a size of the source of deuterons. The measurement is difficult but possible \cite{Graczykowski-Janik-Kisiel}.

\section*{Acknowledgments}

We are grateful to M. Carrington, \L .~Graczykowski, M.~Janik, A.~Kisiel and W.~Rz\c esa for fruitful discussions. This work was partially supported by the National Science Centre, Poland under grant 2018/29/B/ST2/00646.

\end{document}